# Role of Single Chemical Heterogeneities in Generating Anisotropic Tactile Sensitivity and Soft Sliding Friction Phenomena

Kayla A. Hepler[1], Leanne Ton[1], and Charles B. Dhong[1,2]*

[1] Department of Materials Science and Engineering, University of Delaware, Newark, DE 19716

[2] Department of Biomedical Engineering, University of Delaware, Newark, DE 19716

*Author to whom correspondence should be addressed: cdhong@udel.edu

## Abstract

Physical heterogeneities in the context of sliding friction—such as a human finger exploring an object—have been well studied, yet the behavior of chemical heterogeneities in mesoscale soft sliding remains underexplored, despite the similar prevalence of chemical and physical variations in real systems. Here, we experimentally characterized the friction of a planar soft elastic probe sliding across a single chemical heterogeneity that was formed at the interface of two silanes on silicon wafers. By constructing phase maps across multiple loads and velocities, we quantified the occurrence of several frictional phenomena at and around the chemical edge, including stiction spike formation, edge slope direction, baseline shifts, and baseline drift, and quantified their sliding direction-dependent formation. We found that chemical heterogeneities made by more disparate materials (butyl- and aminopropyl-terminated) exhibited several phenomena that were more often direction-independent compared to chemical heterogeneities formed from more similar materials (butyl- and hexyl-terminated). We attributed this directional asymmetry to elastic body effects. In subsequent human testing (n=36), we observed that humans also exhibited directional-dependent accuracy (66.7% versus 38.9%) on one pair (butyl- and hexyl-terminated) but not the other (77.8% versus 75%), which in the context of our phase maps, suggests that the slope of the friction force when sliding over a chemical edge is important for generating a clear edge of a tactile feature, rather than the differences in simple material properties or other friction phenomena.

## Introduction

Soft sliding friction is widespread in engineering and biology, such as rubber tires sliding on pavement[1], geckos climbing on walls[2], and contacts in the eye[3]. Of particular interest in this study are human fingertips, which produce friction when swiping against an object[4,5]. The introductory view of sliding friction in rigid objects[6–10], which are dominated by asperity interactions and plastic deformation, differs significantly if the sliding probe is soft and deformable because deformability introduces or amplifies factors like energy dissipation, adhesion, and interfacial instabilities such as Schallamach waves.[11–18]

In addition to accounting for deformability in soft objects, real surfaces also contain chemical and physical heterogeneities, which originate either from inherent variation, or can be introduced intentionally to engineer friction properties, like anisotropy or increased performance in the bioinspired literature on geckos[19–21]. Here, we were interested in how sliding across two different surface chemistries impacts the perception of touch for applications in tactile aids. Most studies in friction have predominately focused on physical heterogeneities from surface roughness, especially on the multiscale impact of surface asperities[18,22–24] and even in the context of haptics or human perception[15,25–28]. A fewer number of studies are focused on patterned or well-defined physical heterogeneities, i.e., a single edge. On the macroscale, Janko et al. showed that friction of the fingertip over a single physical edge is governed by contact geometry[29]. They found that step-up features (sliding from lower to abruptly higher height) increased friction due to compression at the leading edge, while step-down (sliding from the higher side down to the lower side) transitions had smaller forces due to less deformation of the fingertip. On the nanoscale, frictional transitions over step edges have been investigated using atomic force microscopy. Chen et. al found that on atomic graphene step edges, step-up transitions produce pronounced resistive forces, while step-down transitions can exhibit both resistive and assistive components due to a combination of topographic and chemical interactions[30]. Similar behavior has been observed on ionic crystal surfaces where step-down motion can become assistive as the tip relaxes into a lower-energy region of the interaction potential[31]. Despite physical heterogeneity being studied at both the nano and macroscale, including in the context of human touch, it remains unclear how a chemical heterogeneity impacts soft sliding friction at the macroscale. It is unclear because many of the mechanisms like the (e.g., forced loss of contact over a feature) and geometric arguments (e.g., physical collision into a barrier) of physical heterogeneities might not occur with a chemical heterogeneity.

Well-defined studies of chemical heterogeneities at the nanoscale have demonstrated that friction and adhesion can vary significantly across surfaces with controlled chemical contrast, even in the absence of measurable topography. Chemical force microscopy and lateral force microscopy have been used to probe self-assembled monolayers, revealing distinct friction and adhesion responses associated with different terminal functional groups[32–34]. Similarly, adhesion has been shown to vary across chemically heterogeneous interfaces in controlled systems, including micropatterned charge domains created through surface functionalization[35]. While these studies provide clear evidence that chemistry alone can modulate interfacial forces, they are largely confined to highly controlled, spatially well-defined nanoscale environments and do not fully capture the behavior of heterogeneous chemical interfaces under repeated, macroscale sliding where history-dependent effects become important. That is, in most mechanical characterization studies, friction is measured over repeated, independent sliding events and often reported as an average over multiple trials[36–39]. In contrast, real soft sliding systems such as tires on the road or a human finger exploring a surface experience continuous contact with intermittent breaks which accumulate interfacial and elastic history. As a result, the role of sequential sliding and history-dependent effects and how these are influenced by chemical heterogeneities, in soft systems at the macroscale, remains poorly understood.

Here, we experimentally studied the macroscale friction of a soft rectangular probe slid across a single chemical heterogeneity (edge) formed by two different silanes on a silicon wafer. We repeated experiments at multiple mass and velocity conditions to construct phase maps of the types of frictional phenomena observed, and whether phase maps exhibited symmetry depending on the direction of motion (i.e., sliding from surface A to B, or vice versa). We then found that humans seemed to be sensitive to one of these friction phenomena, even though humans are physiologically capable of sensing the other friction phenomena we identified.

## Experimental Methods & Rationale

**Selection of silanes**. Silanes were chosen as material system due to their commercial availability, relative deposition consistency, and timely ability to coat samples with medium throughput[40]. We have previously used vapor phase deposition of silanes on silicon wafers to investigate human touch in a system where surface chemistry can be varied with minimal differences in surface roughness and physical features[41,42]. Thin silane layers (monolayers or multilayers) are formed through the covalent linkage of the "head" group to the substrate and the "tail" portion of the silane possessing desired functional groups. The presence of true monolayers in silane and self-assembled monolayers (SAM) systems is often not experimentally realized, with studies suggesting that film growth can go beyond a single layer through oligomerization resulting in thin-film or multilayer structures[43–45]. However, regardless of monolayer or multilayer formation, more important to the application here is that the resulting film tends to be within a narrow range of physical variation in the throughput needed for multiple mechanical and human tests.

To create two material pairings, our rationale was to select one pair that can be discriminated by humans and the other is not discriminable. Three materials of n-butyltrichlorosilane (C4), n-hexyltrichlorosilane (C6), and (3-aminopropyl)trimethoxysilane (APTMS) were chosen. In our previous work[41], the pairing of C4 versus APTMS demonstrated high discriminability in 3 alternative forced choice (3 AFC) human psychophysical testing due to APTMS having a terminal amine increasing surface energy and molecular ordering as compared to C4. Conversely, C4 versus C6 was chosen as the second material pairing due to their similarity and likely difficulty for humans to discriminate. Conveniently, C4 is present in both pairings.

**Sample preparation.** The silanes used in this study were purchased from Gelest and Ambeed (95-100% purity) and used as received. A glass vacuum desiccator was cleaned with acetone and ethanol, dried with nitrogen, and preheated at 80 °C for one hour. As shown in **Figure 1a**, silicon wafers (University Wafers) were exposed to oxygen plasma (Glow Plasma System, Glow Research) for one minute and then immediately transferred to the preheated desiccator containing ~50 µL of C4 silane on a glass coverslip. The desiccator was evacuated for five minutes and then maintained under static vacuum at 80 °C for one hour.

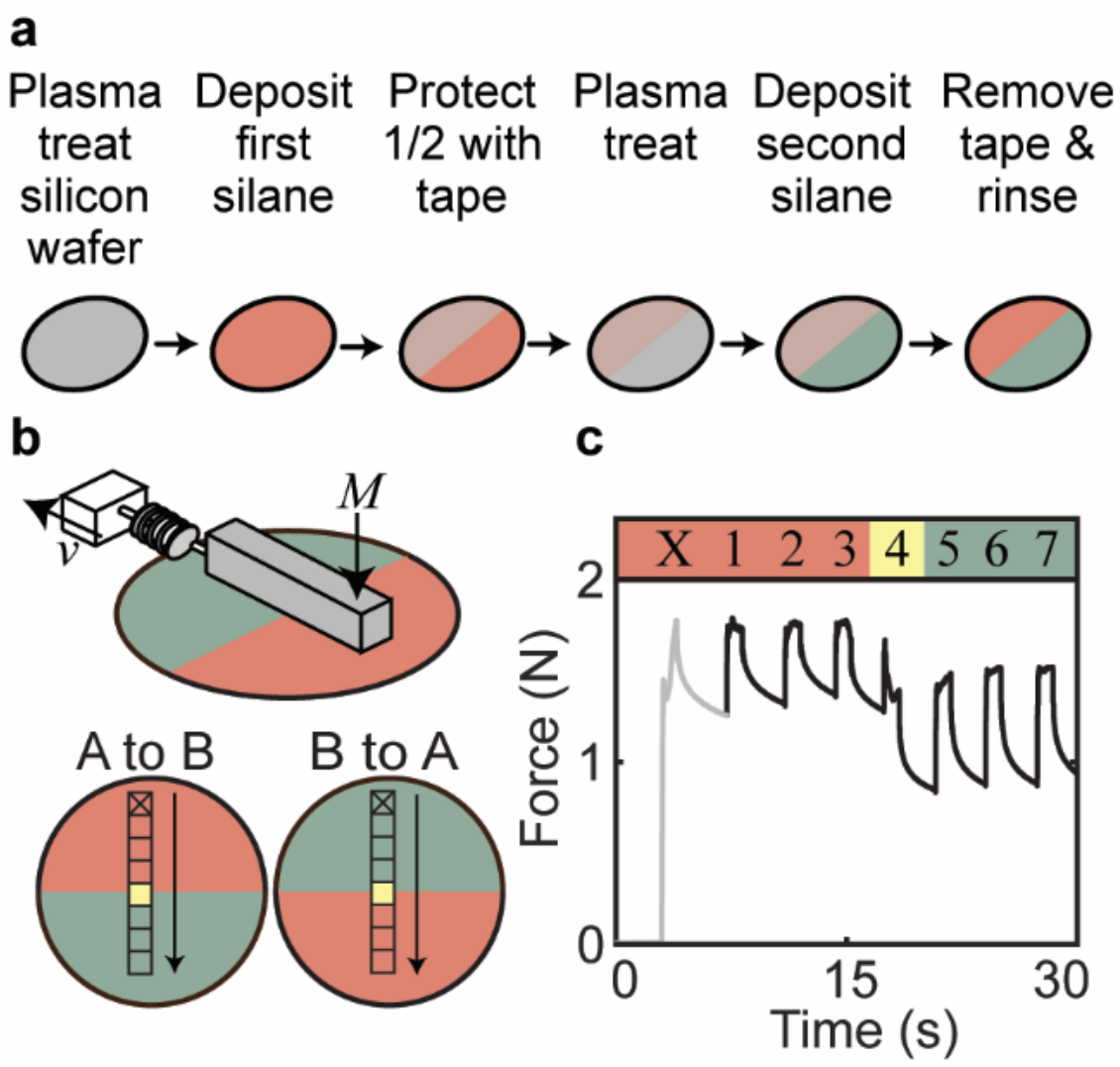

**Figure 1. Schematic of approach overview.** (a) Sample fabrication of silane surface with two silanes. (b) Mechanical testing apparatus with schematic of the mock finger's 8, 10 mm pull locations across the sample surface to collect friction at varying sliding velocities, *v*, and applied mass in addition to the finger, *M*. (c) Representative friction trace showing an initial discard pull (X) followed by three pulls in silane A (1-3) onto a pull at the 'edge' between silanes (4), followed by three silane B pulls (5-7).

Single sided Revalpha thermal release tape (Semiconductor Equipment Corp, 105 °C) was cut in to semicircular masks (50 mm radius) using a Cricut Maker 3. The masks were applied to the C4-coated wafers using a 3D-printed jig to ensure consistent alignment and rolled with a printmaking roller to remove trapped air. The mask boundary was discretely etched on the backside of each wafer to mark its exact location for subsequent measurements.

A second clean glass desiccator was preheated at 80 °C for one hour; separate desiccators were used to prevent potential cross-contamination. The masked wafers were then treated with oxygen plasma for two minutes to remove the previously deposited silane from the unmasked regions. The wafers were then immediately transferred to the preheated desiccator containing ~50 µL of the second silane (C6 or APTMS) on a glass coverslip. The desiccator was evacuated for five minutes and then maintained under static vacuum at 80 °C for one hour.

A hot plate was heated to 110 °C (tape release temperature 105 °C). Wafers were individually placed on the hot plate until the tape detached from the surface (5-10 s), after which the wafers were removed and the tape discarded. Samples were then sonicated for 5 minutes each in toluene, dichloromethane, and ethanol, before drying with nitrogen to remove any residual contaminants from the fabrication process[46]. (see **SI Fig. S2** for AFM verifying residue removal)

**Surface characterization.** Standard surface characterization techniques were used to ensure successful deposition of the silane films and the integrity of the surface after rinsing and exposure to heat release tape.

***X-ray photoelectric spectroscopy (XPS).*** Elemental analysis of silane coated and uncoated silicon wafers was performed via a Thermo Scientific K-Alpha XPS system on 1 cm × 1 cm samples. The spectral data was then analyzed using the Thermo Scientific Avantage System software.

***Atomic force microscopy (AFM).*** A Bruker Multimode AFM was used to characterize the topology and surface roughness of the applied silane coatings. The 1 cm × 1 cm samples were analyzed by tapping mode scans 10 μm × 10 μm area at a rate of 1 Hz and average drive frequency of ~180 kHz. The profiles were then analyzed using Gwyddion software.

***Contact angle hysteresis.***
An Ossila Contact Angle Goniometer was used to measure advancing and receding water contact angles. In the advancing angle measurement, an approximately 4 μl DI water drop was dispensed onto the silane surface. Then, with a syringe attached to micromanipulator, small increments of water were added until the drop spread and an image was taken. For the receding angle, water was slowly extracted until the drop began to retract after which an image was taken. For each surface five drops were measured, and the contact angle of each image was analyzed using an automatic circle fit (ImageJ). The difference between each advancing and receding angle (hysteresis) was also calculated for each measurement pair. The average receding angle, average advancing angle, and standard deviation of the hysteresis is reported.

**Elastic Probe Preparation.** An elastic system with minimal viscoelastic tack was selected. Consistent with prior work[41,47–49], the probe mimics key properties of a human finger (nearly load-independent contact, surface energy, composite modulus, glassy outer layer). The probe consisted of a rectangular prism (1×1×4.8 cm) of polydimethylsiloxane (PDMS, Dow Sylgard 184, 30:1 crosslinker ratio) embedded with an acrylic, 3D printed rod (⌀ = 0.5 cm, $l$ = 4.4 cm). The rod provided mechanical rigidity and mechanical stiffness like that of bone. The PDMS was exposed to UV-ozone for 8 hours to remove tack.

**Mechanical Testing.** Prior to data collection, the etched mask boundary on the backside of each wafer was transferred to the front surface using a permanent marker outside the measurement region. A reference point was marked 45 mm above the mask boundary within the homogeneous region of silane A on either side and connected to define a line. This step accounted for any variation in mask placement, ensuring that friction measurements accurately captured the interface between silanes.

The PDMS elastic probe was connected to a motorized stage (V-508 PIMag Precision Linear Stage, Physik Instrumente) via a 250 g Futek LSB force sensor ($k$ = 13.9 kN m$^{-1}$) and sampled at 560 Hz (Keithley 7510 DMM) to collect traces of the stick-slip friction behavior during sliding. Each sample was placed on the stage, and a low noise vacuum pump was used to hold the wafer in place from below. The elastic probe is mounted at an ~3° angle and lowered until a 1 cm × 1 cm square contact was made with the front edge of the probe aligned to the sample's reference line.

The elastic probe was then slid at 16 total combinations of applied mass ($M = 0, 25, 75, 100$ g, added to the 6 g deadweight of the finger) and velocity ($v = 5, 10, 25, 45$ mm $s^{-1}$) to approximate friction conditions experienced during human exploration. To ensure that we were capturing friction at the chemical heterogeneity, a 1 cm displacement was used due to the 1 $cm^2$ contact area of the elastic probe. The first pull, (marked with a X in the pull numbers across the top of **Fig. 1c**), is discarded to account for any extraneous forces on the mock finger encountered during setup. Without raising the finger, it is then pulled seven times in 1 cm displacements each. The first three pulls occur in the fully homogenous region of silane A (**Fig. 1c**, pink region), followed by a fourth "edge" pull (yellow region) where the probe is in Silane A for 5 mm and Silane B for 5 mm. Finally, three more pulls occur in the fully homogenous region of Silane B (green region). The schematic of this pull paradigm is shown in **Figure 1b** and the sequence was repeated over 3 total regions of the surface in both sliding directions (A onto B and B onto A). Per sliding direction (×2), at each given mass and velocity combination (×16), we obtain data of nine homogenous Silane A pulls, three transition "edge" pulls, and nine homogeneous Silane B pulls.

**Human Psychophysical Testing Additional Details.** To randomize the chemical heterogeneity location the tape was cut into rectangular masks ($h$ = 30-60 mm, $w$ = 100 mm) and aligned by eye to the bottom edge of the wafer; overhanging tape was trimmed with scissors. Following surface preparation, wafers were cut in half (UTILE Precision Wafer and Glass Cutter) and secured into petri dishes using double sided tape. To minimize potential bias from sample geometry, samples were arranged in the Petri dishes as paired halves to visually reconstruct a full wafer, ensuring equal presentation of left and right halves across conditions. Counterbalancing was employed to reduce ordering effects. Three participants were initially trained and tested on the C4/C6 pair, and three on the C4/APTMS pair. Within each material pair, the order of sliding conditions (A onto B and B onto A) was alternated.

After participants were familiar with the sensation, a dot was placed on the center of each participant's index fingernail to standardize finger positioning. After testing, samples were removed, labeled, and the true interface location was transferred to the front surface using permanent marker. For elongated or diagonal marks, the nearest and furthest distances were averaged and rounded to the nearest whole millimeter. The absolute position of the chemical heterogeneity relative to the sample edge was also recorded.

## Results and Discussion

**Materials Characterization of Surfaces.** Successful deposition of n-butyltrichlorosilane (C4), n-hexyltrichlorosilane (C6), and (3-aminopropyl)trimethoxysilane (APTMS) was confirmed by water contact angle, AFM, and XPS. Characterization results are summarized in Table 1 (raw data in **Fig. S1** of **SI**), with trends consistent to prior work[41,42]: APTMS had the lowest contact angle due to its terminal amine and hydrophobicity moderately increases with chain length from C4 to C6. The roughness ($S_a$) and Hurst exponent were obtained by AFM. Overall, characterization would suggest that APTMS could be informally referred to a "higher friction" surface than C4 due to a higher surface energy (lower water contact angle). C4 is largely similar to C6 by these characterization techniques.

Analysis of the power spectral density (PSD, SI) shows that APTMS exhibits consistent behavior across length scales, while C4 and C6 show a change in scaling behavior at smaller length scales (higher spatial frequencies). Hurst exponent was determined from the high spatial frequency region of the PSD to capture nanoscale surface behavior. The Hurst exponent for C4 ($H \approx 0.44$) is lower than previously observed for pure C4 depositions[41,42], indicating increased nanoscale surface variation, which may result from the thermal release tape masking and solvent cleaning steps used to fabricate dual-silane interfaces. To maintain consistency, C4 was designated as the masked region in all samples.

Despite these differences, roughness values ($S_a$) across all samples remain low, with differences less than 150 pm, below the threshold of human perceptibility[27]. Additionally, low contact angle hysteresis indicates minimal surface heterogeneity. Taken together, these results support that physical contributions to friction are limited, allowing chemical contributions to dominate the observed responses.

**Table 1. Silane precursors and surface characterization**

| Silane Precursor | Surface Structure | Water contact angle (°) | Roughness, $S_a$ (pm) | Hurst exponent, H |
|---|---|---|---|---|
| n-Aminopropyltrimethoxysilane **(APTMS)** | $NH_2$ | (45.3–50.2) ± 1.2 | 784 | 0.14 |
| n-Butyltrichlorosilane **(C4)** | | (87.6–94.1) ± 1.2 | 660 | 0.44 |
| n-Hexyltrichlorosilane **(C6)** | | (101.1–104.4) ± 1.8 | 796 | 0.58 |

**Mesoscale friction.** A custom setup with a planar elastic probe with a rigid cylindrical internal support (1 cm × 1 cm nominal contact area, details in **Methods**) was used to collect sliding friction measurements at a series of applied loads ($M$ = 0, 25, 75 100 g, 0 is a deadweight of the ~6 g probe) and sliding velocities ($v$ = 5, 10, 25, 45 mm/s) relevant to human movement during tactile exploration. Each experiment was conducted in triplicate. Representative friction traces at $M$ = 25 g, $v$ = 5 mm/s for sliding from APTMS onto C4 and the reverse, from C4 onto APTMS (Set 1) are shown in **Figure 2a** whereas C4 onto C6 and C6 onto C4 (Set 2) are shown in **Figure 2b**.

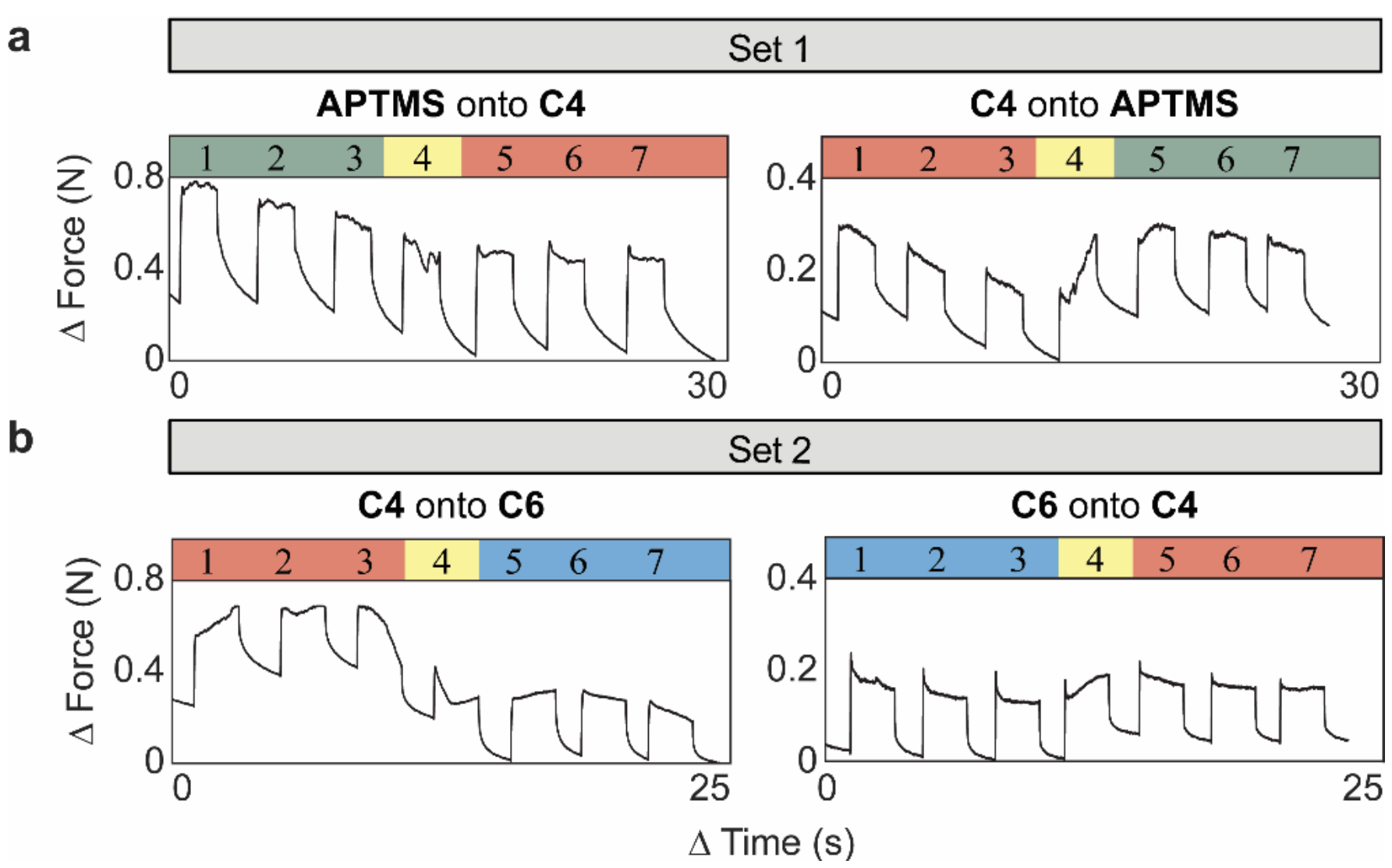

**Figure 2. Representative sliding friction traces.** Friction traces of a mock finger slid across a wafer containing two regions with different silanes, at a velocity of 5 mm/s and an applied mass of 25 g. (a) In the top left, a mock finger began sliding on APTMS, *green*, for pulls 1-3, and then slid on C4, *red*, for pulls 5-7, and the pull over the interface between the two surfaces (the "edge") is indicated in *yellow*. The top right represents the reverse direction: starting on C4 and then sliding onto APTMS. (b) Mock finger slid across Set 2 consisting of C4 and C6. Bottom left represents starting in C4, *red*, and ending in C6 (*blue*), and bottom right represents starting in C6 and ending in C4. Data shows one set of results out of the 16 combinations of velocity (5-45 mm/s) and mass (0-100 g, added to deadweight of mock finger). Each condition was run in triplicate.

In **Figure 2 left column**, we see that sliding onto a second surface showed a decrease in baseline force after the edge and a negative slope over the edge. Reversing the order of the starting and ending surfaces (right column), we observe increase in the baseline force after the edge and a positive slope over the edge. This symmetric behavior is expected from the informal view as a higher surface energy leading to a higher friction surface (APTMS > C4 > C6). However, while the increase or decrease (sign of the change) shows consistency and symmetry, the magnitude of force change is not equal in both directions and thus dependent on the starting surface. We can see this as, in **Figure 2a**, starting on APTMS, the friction of sliding on APTMS goes to a value of 0.8 N, for a change of ~0.4 N over the baseline. Yet, when starting on C4, sliding on APTMS, the APTMS goes to a value of 0.3 N, for a change of ~0.2 N over the baseline. Complicating the picture, the direction-dependent increase and corresponding symmetrical decrease in the other direction is not observed under some masses and velocities, indicating a need to evaluate the behavior of this friction phenomena of a shifting baseline, potentially due to a mechanical history of sliding over an edge. (These are the subject of **Fig. 4-6**).

Overall, we observed several different phenomena that occur in the pull that dictate the shape of the edge (pull 4), and in the pulls across the homogenous regions before (pulls 1-3) and after (pulls 5-7) edge. Across multiple conditions, we see that these frictional phenomena depend on the composition, the order of surfaces encountered during sliding, and the applied load and velocity. The phase maps of where certain frictional phenomena occur, their direction-dependent symmetry or lack thereof, and the potential mechanisms responsible is explored in the next section.

**Prevalence of stiction spike formation in pulls over the 'edge'.** One phenomenon we observed was that the friction trace at the edge sometimes, but not always, generated a stiction spike (**Fig. 3a**; green numbers indicate the number of trials out of three that produced a spike, with 3 indicating all three trials generated a stiction spike). For all edge pulls, the finger was positioned such that it began in contact with both surfaces simultaneously at the start of the 1 cm pull, meaning the last homogeneous pull before the edge always ended with the finger having already contacted a small portion of the second surface. In Set 2, stiction spikes occurred consistently at high mass and high velocity regardless of sliding direction, i.e., whether the mock finger started on C4 and slid onto C6 or vice versa. In Set 1, however, stiction spike generation was both direction-dependent and less frequent. C4 onto APTMS produced fewer spikes than APTMS onto C4, and the conditions under which spikes occurred differed between directions. This is notable because C4 and APTMS are chemically more distinctive than C4 and C6, as APTMS carries a terminal amine capable of hydrogen bonding, yet Set 1 produced fewer stiction spikes than Set 2.

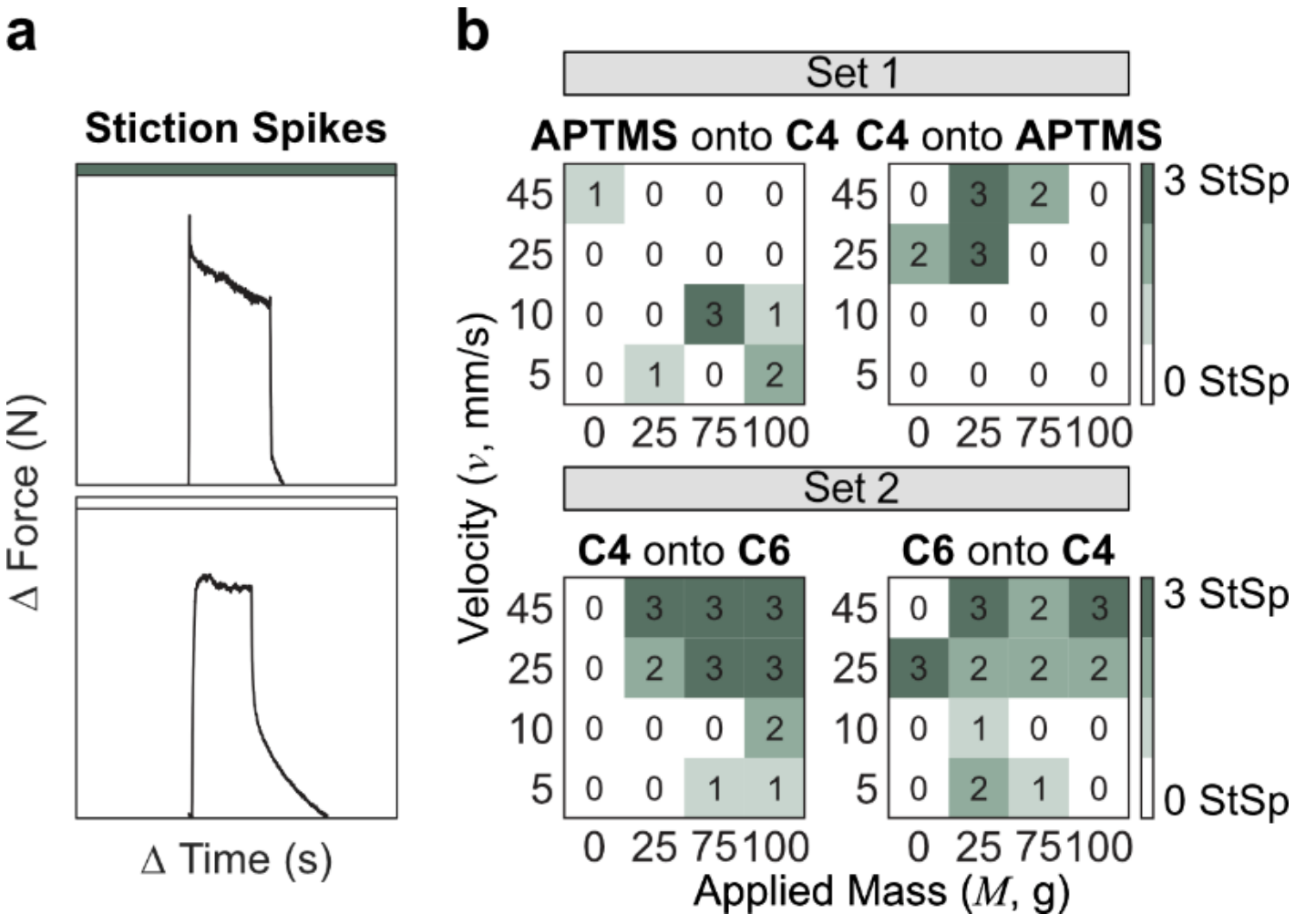

**Figure 3. Stiction spikes of edge pulls.** (a) Representative example of the edge pull containing a stiction spike, *(green)* or not *(white)*. (b) phase maps of the edge pull's propensity to form a stiction spike at all mass and velocity combinations for both directions of the two material pairings. (c) table showing the overall percentage of time the edge pull generated a stiction spike.

The greater frequency in stiction spike generation in Set 2 despite lower chemical contrast between C4 and C6 suggests that surface energy difference alone does not drive this behavior. We identified two possible mechanisms: that the hybrid contact of C4 and C6 together forms a combined interface suitable for stiction spike generation, or that C4 and C6 in their homogeneous regions are sliding in region more prone to generating a stiction spike upon encountering a perturbation like an edge. The observation that C4 sliding onto APTMS also tends to generate stiction spikes at high velocity supports this interpretation, as it suggests that the propensity for stiction spike generation is due to the C4 sliding behavior prior to the edge, as opposed to the chemical contrast generated between surfaces.

**Slope of friction force when sliding over the edge.** In addition to stiction spikes, we analyzed the slope of the friction force when sliding over the edge (**Fig. 4**). We parameterized the force of the edge pull for each of the three trials into "negative" (-1, Blue), "positive" (+1, Red), or "flat" (+0, White) categories by constructing a slope from the points at the end and beginning of the edge pull or, if present, just after the stiction spike with a threshold of 0.01 N. Representative friction of the edge pull for each category is seen in **Figure 4a**. (A maximum value of three or negative three in **Fig. 4b** corresponds to all three trails having the same slope direction.)

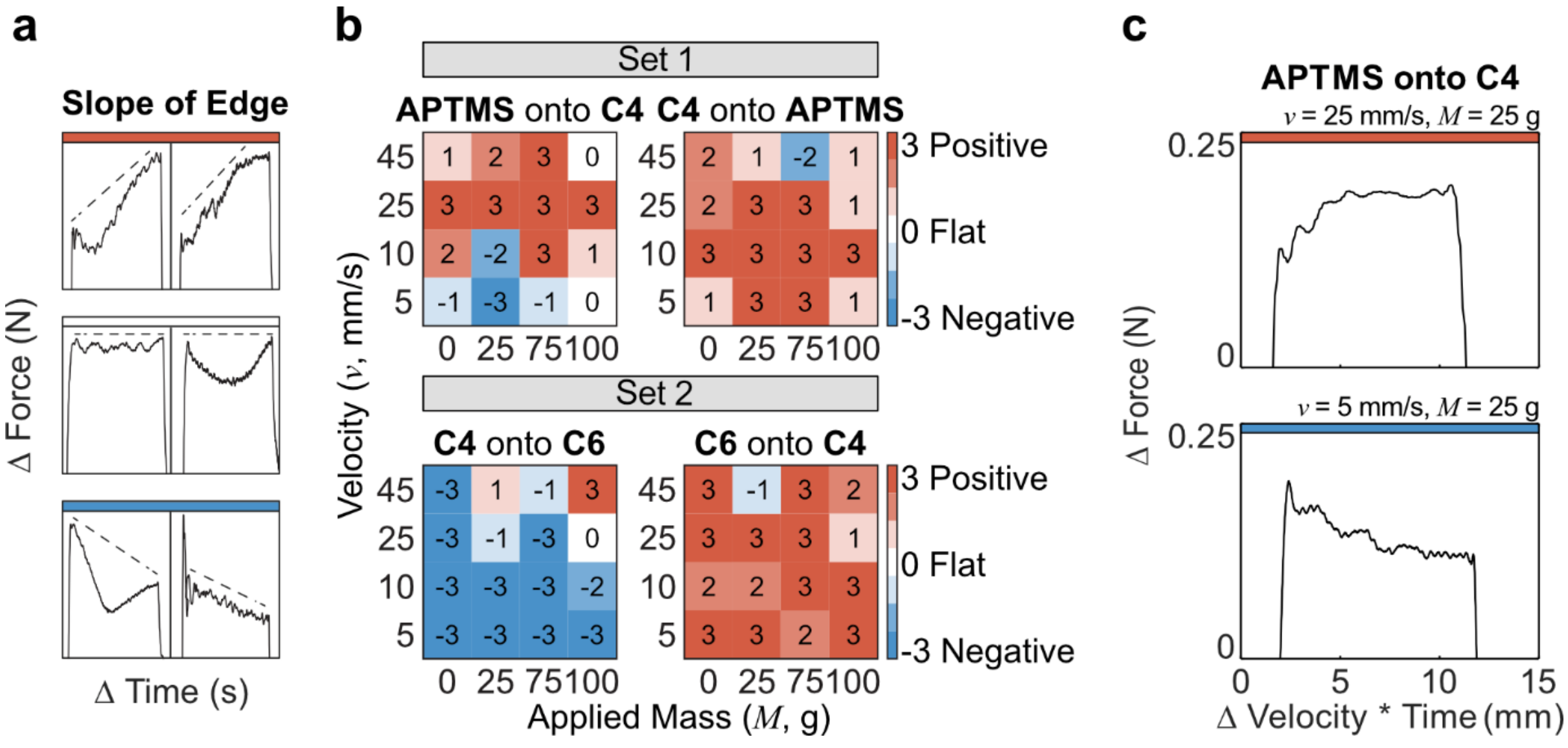

**Figure 4. Slope of edge pull.** (a) Representative examples of the edge pull having positive (red), flat (white), and negative (blue) slope. (b) phase maps of the edge pull's slope at all mass and velocity combinations in both sliding directions of the two material pairings. (c) Example friction traces from red and blue regions of the APTMS onto C4 phase map. Top (red) showing two distinct slopes created during a single pull at higher velocity and bottom (blue) showing a single negative slope at a lower velocity.

A simplified view would suggest that sliding onto a surface with higher friction relative to the first homogenous region would correlate to the friction force over the edge generating a positive slope. Similarly, in reverse, sliding onto a lower friction region would result in a negative slope. **Figure 4b, bottom row**, shows this simplified view is largely consistent in Set 2: sliding onto C6, which is lower friction relative to C4 has a negative slope over the edge whereas sliding onto C4, higher in friction has a positive slope over the edge. We see some deviation from this simplified view in Set 2 at high masses and velocities. Set 1 does not show the same symmetry. Instead, both sliding directions, APTMS onto C4 and C4 onto APTMS, are dominated by positive slopes. Under the previously mentioned "simplified view", APTMS onto C4 would be expected to produce a negative (blue) slope because C4 is lower friction relative to APTMS. The observed positive slope therefore indicates a discrepancy between the expected friction decrease and the experiments. We do note that regions of negative slope are still observed at lower velocities, suggesting that this deviation from the "simplified view" is velocity dependent.

**Figure 4c** shows one possible mechanism for the asymmetry of why APTMS onto C4 still largely yields positive slopes as a function of velocity. At higher velocities, the sloped region (shoulder) after initiation of motion indicates that the mock finger is likely pinned on the higher friction APTMS region as it crosses the boundary, forming reduced contact in the C4 region. The resulting rising force counteracts the expected drop onto the lower friction surface. As the system transitions from a pinned state at APTMS to a detached configuration, contact in the C4 region becomes more uniform and steady sliding commences. This marks a transition to a plateau in the measured response, reflecting re-established contact on C4.

**Baseline shifts in homogenous regions after the edge.** Beyond the friction force taken at the edge, we also observed and classified shifts in the baseline friction force in the homogenous regions before and after sliding over the edge. The baseline shift between the two homogenous regions was quantified by quantifying the baseline friction force before and after the edge. (**Fig. 5a**, baseline before calculated as by averaging the force at the start of pull III and pull IV, baseline after calculated by averaging the force at the start of pull V and VI). By comparing the baseline of the pull before and after, each trial was quantified as a "step down" (-1), "step up" (+1), or "no shift" (+0), with "no shift" defined as a baseline shift smaller than ± 0.01N.

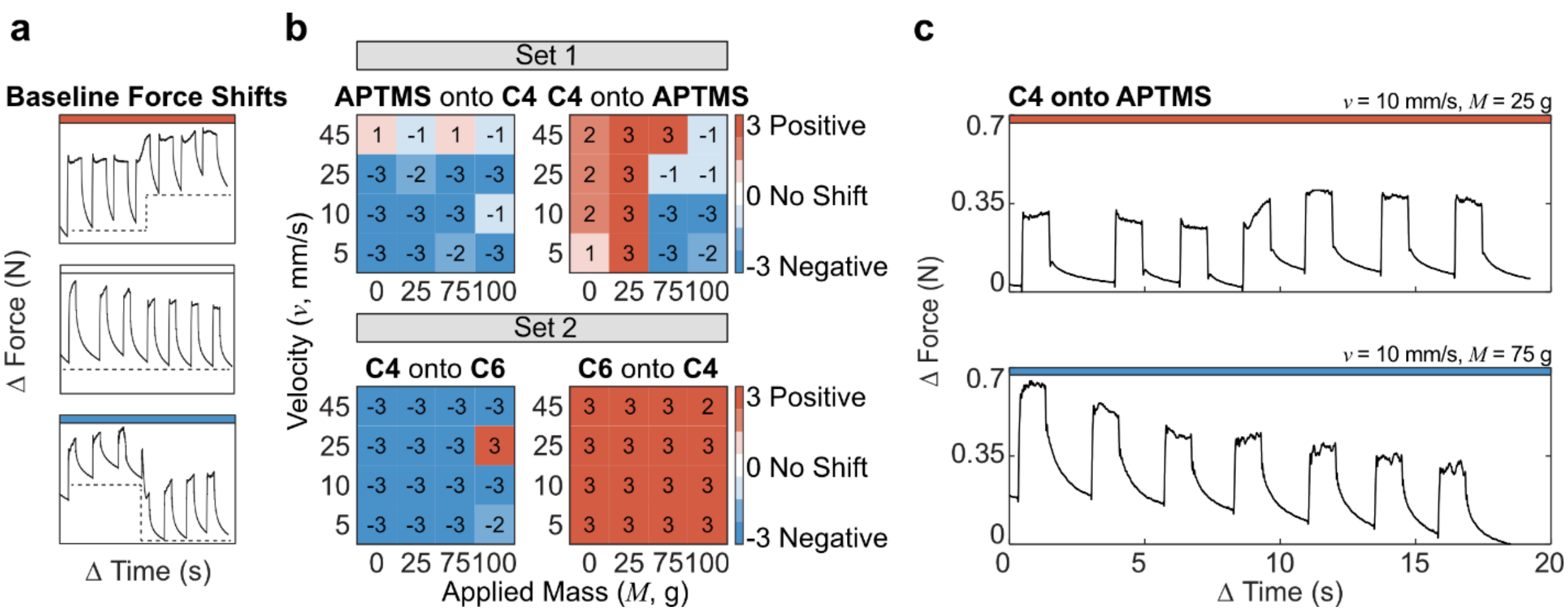

**Figure 5. Baseline shift in force after edge.** (a) Representative examples of baseline force values increasing, *(red)*, staying the same, *(white)*, decreasing, *(blue)* after the finger encounters the edge. (b) phase maps of the baseline force behavior after the edge at all mass and velocity combinations in both sliding directions of the two material pairings. (c) Example friction traces from red and blue regions of the C4 onto APTMS phase map. Top (red)

showing a “step-up” in the baseline force after the edge at lower applied mass (25 g) and bottom (blue) a “step-down” in baseline force after the edge at higher applied mass (75 g).

In many of the conditions in **Figure 5b**, the we see the same expected trend from the previous section on Slope of Edge: sliding onto a higher friction surface leads to a “step up” (higher force) baseline, whereas sliding onto a lower friction surface leads to a “step down” (lower friction force) after the edge. Set 2 follows this trend and is symmetrical, with the only deviation from expected behavior when sliding onto C6 from C4 at $M = 100$ g, $v = 5$ mm/s. We again see deviation in expected symmetry with Set 1 as observed in the Slope of Edge analysis. However, unlike the Slope of Edge analysis, the deviation in expected symmetry was observed in the scenario of sliding from C4 to APTMS. More specifically, we would expect C4 onto APTMS to have “step up” after the edge, which we observe at lower masses. However, at higher applied loads we see “step down”, which was unexpected, and in this case, indicates load-dependent anisotropy.

While a positive shift when sliding onto a higher friction surface is intuitive, a negative shift onto what is otherwise a higher friction surface is not intuitive. **Figure 5c** depicts one possible mechanism for the load-based anisotropy. Most visible in last three pulls, at higher loads ($M$=75 g, lower panel), slow frictional waves have a higher in amplitude than lower applied mass ($M$=25 g, upper panel). These larger frictional waves likely mean a larger loss of contact area that may not be fully reforming between different sequential pulls. Thus, even though the friction force per unit area may be higher in APTMS, the lower contact area is yielding a lower overall friction.

**Baseline drift in homogenous regions occurring before and after the edge.** In some cases, the friction force exhibited a baseline drift between pulls in the homogenous regions (**Fig. 6a**). While initially considered experimental error, the drift was observed consistently in repeated measurements and not observed at all velocities and normal loads. Furthermore, both positive and negative drift were observed, yet the sign rarely switched per homogenous region. Finally, one material, C6 (**Fig. 6b bottom right**), rarely showed drift whether it was before or after the edge and every material showed at least a few regions where there was no drift. We could not attribute this due to an artifact of a clustered series of sample testing on the mechanical testing apparatus as we had intentionally semi-randomized our testing schedule and parameter exploration.

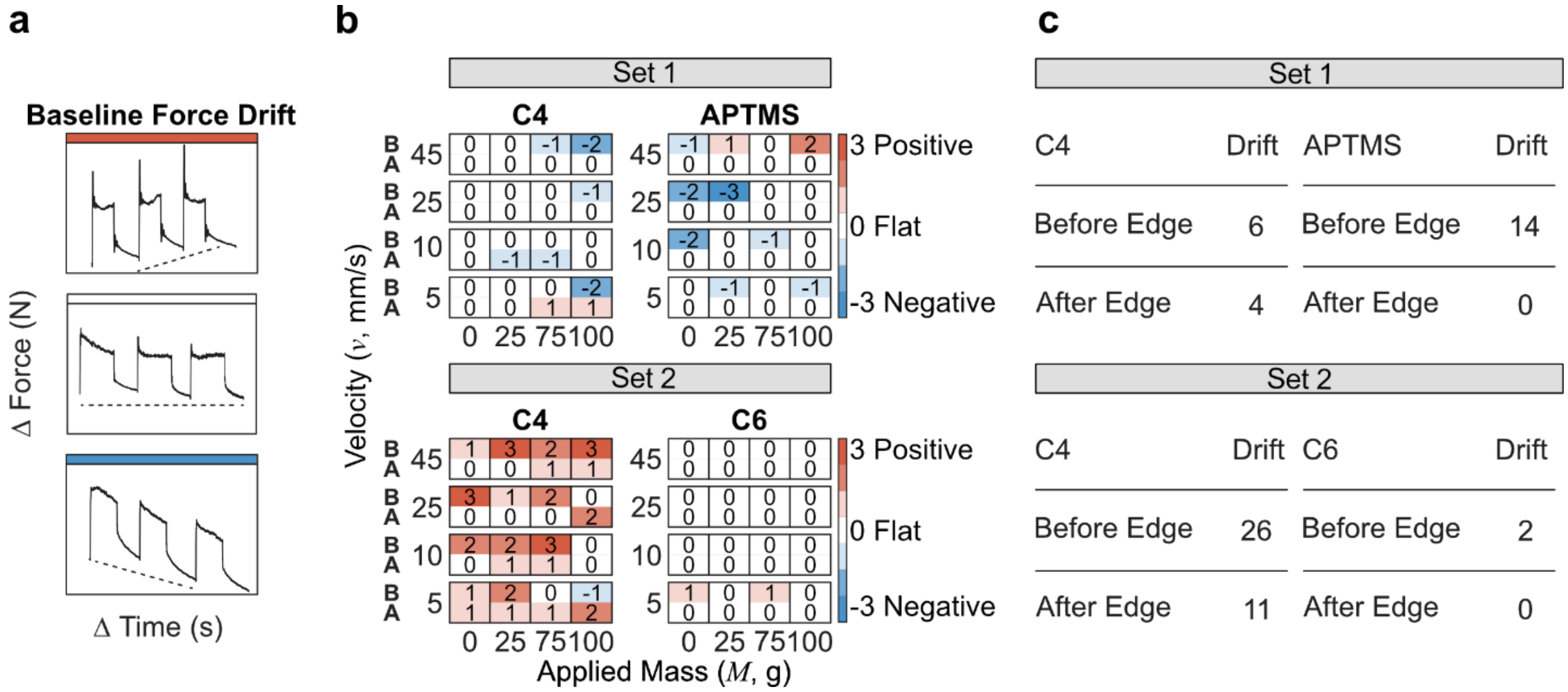

**Figure 6. Baseline drift before and after edge.** (a) Representative examples of baseline drift values increasing, *(red)*, staying the same, *(white)*, and decreasing, *(blue)* in the homogenous regions. (b) phase maps of the baseline drift behavior for each material in Set 1 and Set 2 before (**B**) and after (**A**) the edge. (c) Total instances of drift for each material when it was before and after the edge in Set 1 and Set 2.

The drift from each of the three trials was quantified onto "negative" (-1), "positive" (+1), or "flat" (+0) (**Fig. 6a**) by constructing a slope from the baseline point just prior to the first pull to the baseline point just prior to the last pull. This was performed in both homogeneous regions: Before the edge ("B", pulls 1-3) and After the edge ("A", pulls 5-7), and a maximum value of three or negative three corresponds to all three trials resulting in the same direction of drift (**Fig. 6b**). We set a threshold of 0.1 N by calibrating against C4, which was present in both Set 1 and Set 2 and thus we can expect that the drift in C4 Before the edge should be minimized. Thus, we expect any trends that are visible above this threshold to be due to phenomena specific to sliding friction and not purely experimental variation or errors.

We see that C4 (**Fig. 6b left column**) still shows several areas with drift Before the edge, despite setting a moderately high threshold of 0.1 N. Thus, one explanation for the difference in Before the edge of C4 in Set 1 and Set 2 is a baseline level of variation that likely due to some experimental setup, or difficult-to-control parameters. However, it may have been unfortunate that C4 was the common material in both Sets because this drift in C4 may be inherent to C4 and material dependent. C6 (**Fig. 6b bottom right**) almost never shows drift, with a few examples of drift occurring when the finger is slid on C6 first, and no examples of drift occurring when sliding onto C6. Supporting the material-dependence and the impact of traversing an edge on drift, in APTMS Before the edge, the results consistently show negative drift at many velocities and masses (**Fig. 6b top right**). We note that a threshold of 0.1 N indicates a robust effect from drift. In APTMS, 0.1 N is almost comparable in magnitude to the total signal. However, when APTMS is After the edge, we do not see drift even though we observed APTMS Before the edge generates significant drift. In contrast, by comparing both C4 After the edge in Set 1 and Set 2, we see that the drift is dependent on what the finger was sliding on prior to the C4.

Initially, we were surprised that C4 Before the edge in Set 1 and Set 2 were not identical. However, we see that drift is consistent, can be positive or negative, and can be suppressed consistently in other materials. Therefore, one explanation is that C4 is a comparatively variable surface, likely because it can generate several different instabilities during its sliding at these masses and velocities. We observe the drift is material and condition dependent, with the edge acting in some cases in a binary manner: the existence of an edge may allow for stress relaxation in the mock finger which may eliminate drift. Furthermore, given that drift occurs in both positive and negative directions, and was not limited to certain pull numbers, this drift is not fully explained by geometric effects from the small angle of the mock finger. We also observed that APTMS no longer generates drift if it is the surface after the edge, despite consistently generating drift Before the edge. For the samples here, going over an edge seems to reduce drift (**Fig. 6c**) because there is more drift, and more variability in the direction of drift, before an edge than after the edge. We first believed drift was purely from experimental error, but given its material dependence, this suggests that baseline drifts in the interpretation and analysis of soft sliding friction may need to be more widely considered.

**Human Detection of the Edge by Psychophysical Testing.**

To probe which of the discussed frictional phenomena might be important for human touch perception, with applications in improving tactile devices, we performed human psychophysical testing (IRB# 1484385). Participants (*N*=6) were asked to find the chemical edge, and the location of the chemical edge was randomized by varying the placement of the thermal release tape during fabrication. (**Fig. 7**) Prior to the task, participants were presented with a single Petri dish containing an example for each sliding direction (sample one: Surface A onto Surface B and sample 2: Surface B onto Surface A) for familiarization. In these training samples, the location of the chemical edge was indicated with a small permanent marker line along the periphery. After familiarization, participants were given unmarked samples with a chemical edge that varied in location. Participants were asked to find the edge within 30 s and mark it with a fine-point tip (see **Methods**). Each Set was tested 12 times (6 per direction). The distance between the participant's mark and the true interface was measured. Both familiarization and testing were counterbalanced to reduce order effects. The apparent contact area of the fingertip is much larger than the size of the chemical heterogeneity ($\geq 10$ mm$^2$)[50]. Thus, participant responses were classified as "Correct" when the marked location was within 10 mm of the chemical edge. **Figure 7b** shows results from testing.

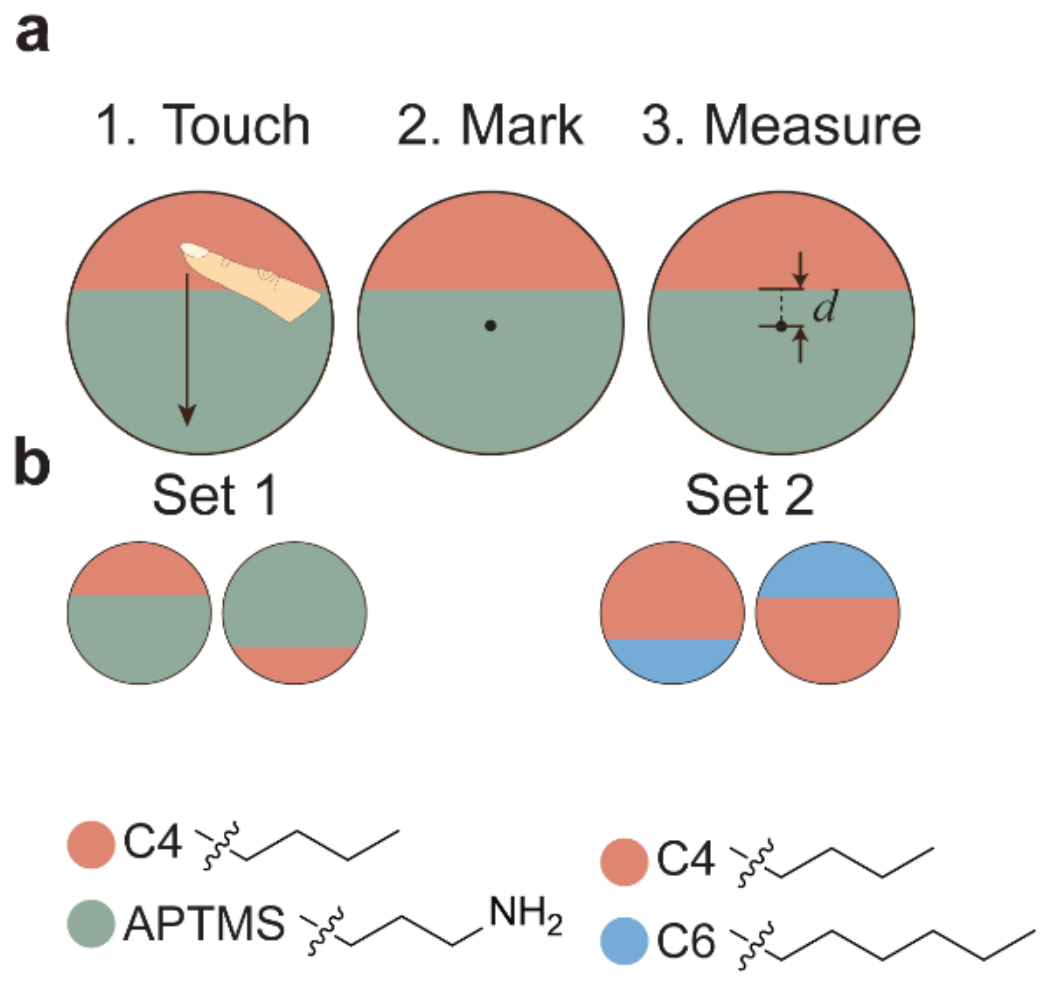

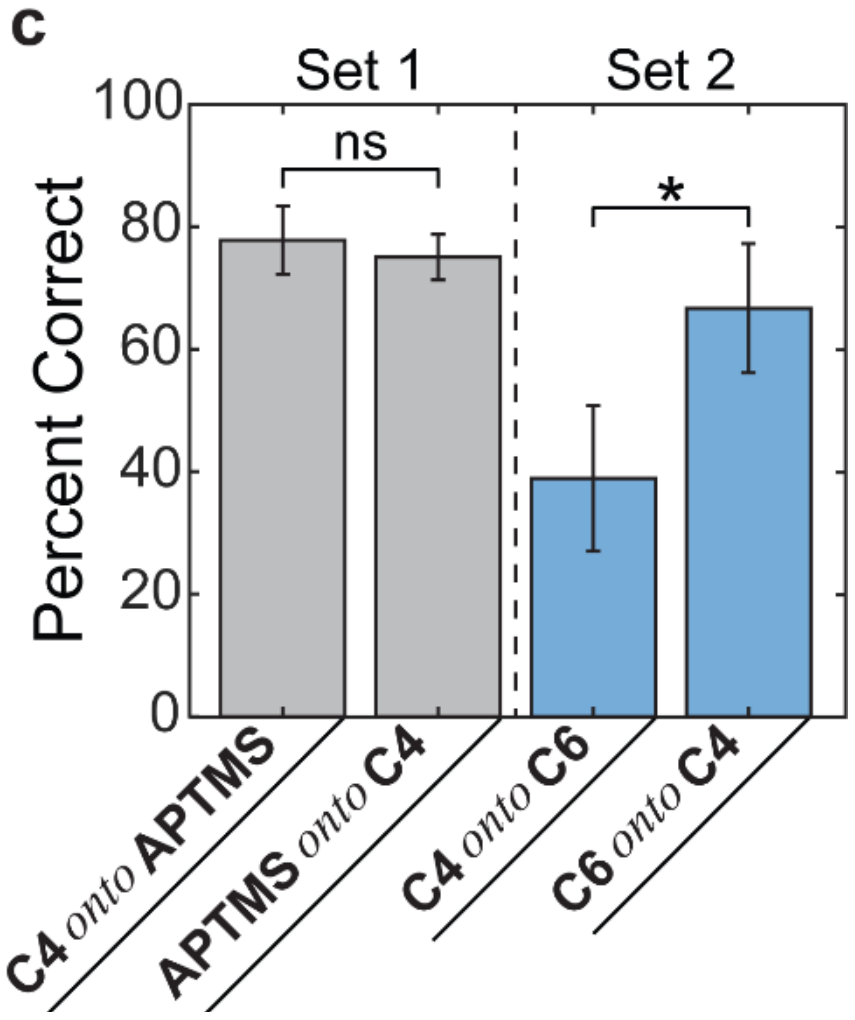

**Figure 7. Human psychophysical testing results.** (a) Schematic of the psychophysical task in which participants slid their fingers from the first silane onto the second, marked the perceived location of the chemical heterogeneity, and the distance, *d,* to true heterogeneity location measured. (b) Schematic showing reversal of sliding direction, variation of 'edge' location, and silane chemistries for Sets 1 and 2**.** (c) Results of psychophysical testing with human subjects. Error bars indicate ± standard error of the mean. Asterisks represent significance ($p < 0.05$, Fisher's exact test). Each condition includes $n = 36$ trials across $N = 6$ subjects.

In Set 1, C4 onto APTMS and APTMS onto C4 had similar success rate that were independent of the sliding direction (77.8% and 75.0%, respectively). However, in Set 2, directional anisotropy was observed. For C6 onto C4 participants correctly identified the "edge" 66.7% of the time, whereas in the reverse direction, C4 onto C6, performance decreased to 38.9%. Consistent with this trend, the average deviation from the true interface for C4 onto C6 was ~13 mm, compared to ~7 mm for the other conditions.

To understand the source of this anisotropy, these results were compared to the mesoscale friction phase maps. The phenomena where C4 onto C6 is unique and unlike the others (C6 onto C4, APTMS onto C4, C4 onto APTMS) is the slope of the friction force across the edge. Specifically, the friction force decreases across the edge from traversing from C4 onto C6, whereas all other conditions exhibit an increase. We note that the friction phase maps were constructed from multiple short, controlled pulls, whereas participants identified the interface through unconstrained exploration, typically beginning with longer continuous strokes followed by shorter confirmatory motions. Despite this difference in interaction, the elastic probe phenomena is likely characterizing key trends that are reflected in human testing results.

## Discussion

Throughout the classifications of friction phenomena in this study, Set 2 was consistently more symmetric than Set 1. For example, for Set 2 (**Fig. 4b),** the slope of the edge for C4 onto C6 was predominately negative (blue) whereas the reverse of C6 onto C4 was predominately positive (red). Set 2 was also more consistent and had fewer individual exceptions in the phase maps. In contrast, Set 1 was not symmetric: in the same **Figure 4b**, the slope of the edge was always majority positive (red) regardless of sliding from C4 onto APTMS or sliding from APTMS onto C4. For both Sets, we attributed instances of asymmetries to elastic body effects.

The widespread lack of symmetry in Set 1 for the slope of the edge (**Fig. 4c**) was attributed to potential pinning on the APTMS before moving onto C4. For the change in baseline after the edge, (**Fig. 5b**), C4 onto APTMS sometimes showed a negative shift (blue) at high applied loads, and we attributed this to the contact area not being reformed fully between consecutive pulls due to slow friction waves, leading to a lower overall friction force despite APTMS being considered a typically "higher friction" surface than C4. We believe the reason that the contact area is not being reformed in this region of the phase map (i.e., combinations of mass and velocity) because APTMS surfaces tended to form slow frictional waves under these conditions, which we also observed in a prior study[42].

Set 2 was otherwise was mostly symmetric. Most exceptions occurred at high velocity and/or applied mass conditions. These deviations are localized rather than systematic. For example, in C4 onto C6 at 25 mm/s and 100 g, **Figure 4b** shows a neutral (white) slope, whereas most other conditions are negative (blue). In **Figure 5b**, the only positive baseline shift also occurs at 25 mm/s and 100 g. At a slightly higher velocity (45 mm/s, 100 g), a positive slope is observed. We are unsure if these exceptions are purely from instabilities, experimental non-idealities, or other mechanistic rationale, however it does reinforce that sliding friction at the mesoscale can exhibit sharp transitions in behaviors and thus multiple conditions (velocities and masses) should be taken.

There is less prior work on the mesoscale friction of an elastic body over a chemical heterogeneity, where physical differences are minimized. Thus, we draw partial comparisons to friction across physical step-edges. Across length scales, physical step-up transitions produce friction maxima; in soft macroscopic contacts this arises from deformation around the step edge whereas in atomic-scale graphene systems the friction increase is from combined topographic and chemical edge interactions[29,30]. In our chemical heterogeneity system, a qualitatively similar step-up behavior was seen in the force traces, although the mechanism is different. Distinct to chemical edges, there is a rich variety in the change (increase, decrease, or neutral) that is now mass and velocity dependent, indicating potential tunability, whereas a physical step-up would always incur an increase in force at the edge, regardless of condition. Further tuning and decoupling of the forces with chemical edges was observed in APTMS onto C4 where we had a positive slope at the edge, but an overall lower baseline of force when the finger was sliding on the homogenous regions. We suggest one potential mechanism is the pinning on the trailing edge of the elastic body. Further mechanistic differences between physical and chemical edges are that step-down transitions in physical edges reduce friction in soft contacts due to the absence of leading-edge compression, whereas at the nanoscale the frictional response is sensitive to edge chemistry, exhibiting either assistive topography-dominated behavior or resistive forces when the step-edge is chemically functionalized[29,30]. However, the negative for C4 onto C6 in Set 2 is more consistent with a decrease in interfacial shear strength rather than deformation asymmetry or assistive mechanics (slip and recovery of stored elastic energy in the sliding body) like those in a geometric step-down.

**Conclusions**

Soft sliding of an elastic body over a single chemical heterogeneity produces rich frictional behavior with different mechanistic origins than physical heterogeneities, which can produce unique behavior like a direction-independent increase in force over the heterogeneity. By systematically mapping frictional phenomena across load and velocity conditions with two pairs of surface chemistries, and in both sliding directions, we showed that chemical heterogeneities can exhibit direction-dependent formation or suppression of different friction phenomena. Set 1, which had a higher chemical contrast by surface energy and chemical structure, consisting of APTMS and C4 showed more widespread and complex asymmetry than the Set 2, which consisted of C4 and C6 and are more chemically similar. Overall, the broad tunability of chemical edges based on these factors could lead to new strategies for tuning frictional interfaces, including advanced surfaces for micromachines or robotic sensing.

For our intended use in using chemical patterning to generate next-generation tactile aids, human psychophysical testing indicated that participants are sensitive to increases in friction at the chemical edge but not decreases, which connected macroscale friction phenomena to perceptual outcomes that are important in the design of tactile graphics. More broadly, this work demonstrates the value of phase map approaches for characterizing history-dependent friction across a large parameter space and identifies specific conditions that may warrant further investigation through direct contact visualization.[14,16,51]

**Author Contributions**

K.A.H planned and performed experiments, analyzed data, and wrote the manuscript. L.T. performed experiments and analyzed data. C.B.D. planned experiments, analyzed data, and edited the manuscript.

**Conflicts of Interest**

Authors have no conflict of interest to disclose.

**Data Availability**

Data of friction force curves will be provided in a public, online repository at [future location].

**Acknowledgements**

We acknowledge support from the National Eye Institute of the National Institutes of Health under award number R01EY032584. Atomic Force Microscopy was supported by the Delaware INBRE program, with a grant from the National Institute of General Medical Sciences, NIGMS, from the National Institutes of Health and the State of Delaware (P20 GM103446). X-Ray Photoelectron Spectroscopy was conducted at the Surface Analysis Facility at the University of Delaware (NSF CHE-1428149). Studies involving human participants were approved by the Institutional Review Board of the University of Delaware (Project #1484385). Pushpita Bhattacharyya and Jared Medina provided guidance on the design of the human-subject testing protocol. The content is solely the responsibility of the authors and does not necessarily represent the official views of the National Institutes of Health.